# Advancing Nonlinear System Stability Analysis with Hessian Matrix Analysis

Samaneh Alsadat Saeedinia[a], Mojtaba Sharifi[b,*], Seyed Mohammad Hosseindokht[a], Hedieh Jafarpourdavatgar[c]

[a]Department of Electrical Engineering, Iran University of Science and Technology, Tehran, Iran

[b]Department of Mechanical Engineering, San Jose State University, San Jose, California, USA

[c]Department of Electrical Engineering, Amirkabir University of Technology, Tehran, Iran

[*]Corrosponsing Author

**Abstract**

This paper introduces an innovative method for ensuring global stability in a broad array of nonlinear systems. The novel approach enhances the traditional analysis based on Jacobian matrices by incorporating the Taylor series boundary error of estimation and the eigenvalues of the Hessian matrix, resulting in a fresh criterion for global stability. The main strength of this methodology lies in its unrestricted nature regarding the number of equilibrium points or the system's dimension, giving it a competitive edge over alternative methods for global stability analysis. The efficacy of this method has been validated through its application to two established benchmark systems within the industrial domain. The results suggest that the expanded Jacobian stability analysis can ensure global stability under specific circumstances, which are thoroughly elaborated upon in the manuscript. The proposed approach serves as a robust tool for assessing the global stability of nonlinear systems and holds promise for advancing the realms of nonlinear control and optimization.

**Introduction**

In recent years, global stability in nonlinear systems has attracted attention due to their complex dynamics. Various methods have been developed to understand stability in such systems [1,2,3,4,5]. One of the prominent techniques in stability analysis is the Jacobian matrix method, offering insights into local stability[6,7,8]. However, it lacks information on global stability and faces challenges with complex systems[9,10,11].

Researchers are enhancing Jacobian techniques to provide comprehensive insights into system-wide stability. These adaptations aim to overcome limitations and assess global stability in a broader range of nonlinear systems, however these approaches are still restricted to a limited class of nonlinear systems with specific dynamics [6,7,9,10,11].

This paper presents a novel technique for ensuring global stability in a wide range of nonlinear systems. The conventional Jacobian matrix analysis is improved by integrating the Taylor series boundary error of estimation and the Hessian matrix eigenvalues to develop a criterion for global stability in the proposed approach. Its superiority over other methods in global stability analysis stems from its lack of constraints with respect to the number of equilibrium points or the dimension of the system, making it applicable to a broader class of nonlinear systems.

2. **Materials and Methods**

We This paper proposes a methodology to analyse the global stability for n-dimensional ordinary system by analysing Jacobian matrix and Hessian Matrix eigenvalues. To this end, consider an autonomous system as follow:

$$\dot{x}(t) = f(x), \quad x(t_0) = x_0 \in R^n \quad (1)$$

The Jacobian matrix is calculated as below:

$$J(x) = \frac{\partial f(x)}{\partial x} = \left[ \frac{\partial f}{\partial x_1} \quad \cdots \quad \frac{\partial f}{\partial x_n} \right] \quad (2)$$

Eigenvalues of the J(x) are calculated at the equilibrium point.

$$\det(\lambda I_{n \times n} - J) = 0, \quad roots: \quad \lambda_1, \lambda_2, ..., \lambda_n \quad (3)$$

Where det(.)denotes the determinant of a matrix. In this study, we propose a novel methodology for analyzing the global stability of nonlinear systems by introducing a specific condition based on the calculation of error estimation during the linearization process using the Jacobian matrix and Taylor series expansion concept. The goal is to establish a criteria that guarantees global stability for the class of nonlinear systems that satisfy this condition.The proposed approach involves calculating the error resulting from the linearization of the nonlinear system and using it to estimate the accuracy of the linear approximation. By evaluating the eigenvalues of the system solely at its unique equilibrium point, we aim to determine the conditions under which global stability can be ensured.

Specifically, we hypothesize that if all the real parts of the eigenvalues are negative, the nonlinear system is globally stable, subject to the satisfaction of the proposed condition. Conversely, the presence of even one positive eigenvalue indicates system instability. This implies that the system is conditionally stable, depending on the validity of the proposed condition.

This methodology distinguishes itself from conventional Jacobian matrix stability analysis by considering the error estimation of linearized nonlinear systems. By incorporating this error estimation into the stability analysis, we aim to provide a more comprehensive and accurate assessment of global stability.

To validate the proposed approach, we present several examples of nonlinear systems in subsequent sections. Through the numerical solution of the corresponding differential equations and the analysis of state trajectories under various initial conditions, we aim to demonstrate the effectiveness and applicability of our methodology in establishing the criteria for global stability based on error estimation in linearized nonlinear systems.

Consider a general autonomous nonlinear system $\dot{X} = f(X)$ with a unique equilibrium point, denoted as $X_e$. The estimation of *f(X)* can be obtained by calculating the Taylor series expansion and its residual. For a first-degree approximation, the residual is given by equation (4), where R1 is the residual of the first-degree estimation.

$$\dot{X} = f(X) = 0 + J(X_e)(X - X_e) + R_1(X) \quad (4)$$

For an nth-degree approximation, the Lagrange error estimation boundary is defined as shown in equation (5), which $M \geq f^{n+1}(c)$ represents the maximum value of the (n+1)th derivative of f(X) at a specific point $c \in \{[X, X_e] \cup [X_e, X]\}$.

$$|R_n(X)| \leq \frac{M}{(n+1)!}|X - X_e|^{n+1} \quad (5)$$

When employing the Jacobian matrix as a linearizing method, the error of estimation, denoted as Rn, is limited in stability analysis. Specifically, if the estimated function's error is bounded at any point, we can assert that the system is stable against perturbations. Therefore, the residual of the Jacobian matrix-based linearization serves as an indicator of the estimation error and should be bounded for global stability. If the residual is unbounded, it indicates that the error of the first-degree Taylor series approximation is not limited. In such cases, discussing global stability becomes challenging. To achieve global stability criteria, it is essential to establish a bounded boundary for the summation of residuals. By calculating the boundary of the first-degree error estimation and ensuring the existence of a limited boundary for the sum of residuals, we can satisfy the conditions for global stability. This implies that the difference between the original function and its first-degree approximation remains within a certain boundary, allowing us to make conclusive statements about the system's stability. According to the maximum error boundary, the following inequality is held:

$$|f(X)| \leq |(f(X_e) + J(X_e) \cdot (X - X_e)) + R_1| \quad (6)$$

Therefore; if $g(X) = f(X_e) + J(X_e) \cdot (X - X_e) + R_1$ is globally stable, f(X) is globally stable, however, it is important to note that global stability alone is not a sufficient condition to prove asymptotic stability. To establish the spatial conditions that ensure the stability of $g(X)$ and subsequently $f(X)$, this paper introduces a specific class of nonlinear systems, meeting the following conditions:

1) $\text{Real}(eig(J(X_e))) < 0 \quad (7)$

2) $\lim_{X \to +\infty}(X - X_e)^T \max(eig(H(X)))(X - X_e) \leq \varepsilon$

$$(8)$$

Where $\varepsilon > 0$ and H(X) is the Hessian matrix, and the second condition is obtained by considering Lagrange error estimation boundary convergence criteria to a bounded value $\varepsilon$. In the case of $\varepsilon=0$, asymptotic stability can be ensured.

Particular classes of nonlinear systems, which decay rate of their Hessian matrices are faster than $\frac{1}{|X - X_e|^2}$ or have eigenvalues, converging to zero.

Therefore the second criterion can be simplified to the following equation:

$$\max(eig(H(X))) = 0 \quad (9)$$

In the next step, we will evaluate the calculated criteria on two benchmark systems.

## 3. Results

In this section, some applications of the proposed stability analysis method are epitomized and their global stability condition is analyzed by the proposed systemic approach. All the simulation study has been conducted by python 3,10.

### 3.1. Modified FitzHugh-Nagumo model

The FitzHugh-Nagumo model portrays a 2-dimensional nonlinear system that effectively characterizes the dynamic behavior of excitable cells. Explicitly, the model is represented by the subsequent equations:

$$\dot{v} = c(v - \frac{v^3}{3} - w + I - c(v-a))$$
$$\dot{w} = (v + a - bw)/\tau \qquad (10)$$

where $v$ represents the membrane potential, $w$ represents the recovery variable, I is the external current, a and b are constants, c is the feedback gain, and $\tau$ is a time constant. Consider I=0, a=0, b=0.333, c=1 and $\tau = 1$. Phase Portrait of the nonlinear system exhibits global stability, as shown in the figure 1. System has three equilibrium points at (0,0), (3,1), and (-3,-1). Eigenvalues of all three equilibrium points have negative real part, so all the equilibrium points are stable. Hessian matrix of the system is calculated by the following equation:

$$H = \begin{bmatrix} \partial^2 f_1/\partial v^2 & \partial^2 f_1/\partial v \partial w \\ \partial^2 f_2/\partial w \partial v & \partial^2 f_2/\partial w^2 \end{bmatrix} = \begin{bmatrix} -2v & 0 \\ 0 & 0 \end{bmatrix} \quad (11)$$

Since eigenvalues of H(X), vary between $(-\infty, 0]$, $\max(eig(H(X))) = 0$, and consequently the second criteria for global stability is satisfied. Note that maximum eigenvalues for Hessian matrix is located at (0,0), which implies that the main attractor is the equilibrium point at (0,0) and its neighbourhood.

Further analysis is depicted in the phase portrait of the system, see Fig.1.

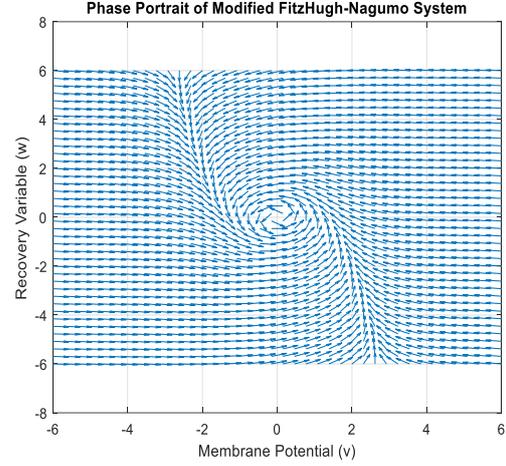

Figure 1 Phase portrait of FitzHugh-Nagmo Model example

Figure (1) reveals global stability of the system.

### 3.2. Van der Pol system

The Van der Pol system equations are:

$$\dot{v} = u$$
$$\dot{u} = \mu(1 - v^2)u - v \qquad (12)$$

where, $v$ indicates the voltage across a capacitor, and $u$ represents the current through an inductor. The parameter $\mu$ controls the nonlinearity of the system. System has a unique equilibrium point at (0,0), and the eigenvalues of the system at the equilibrium point are $\lambda_{1,2} = \frac{\mu}{2} \mp \frac{\sqrt{\mu^2 - 4}}{2}$. To evaluate the proposed methodology, consider $\mu = -0.1$, in which eigenvalues of the system are $\lambda_{1,2} = -0.0500 \mp 0.9987i$, satisfying the first condition (Eq.7). Hessian matrix of the system is $H = \begin{bmatrix} 0 & 0 \\ 2\mu & -4\mu v \end{bmatrix}$. In this case, eigenvalues of H lie on the range of $[0,\infty)$. Therefore, second criterion is not satisfied.

Based on Equation 12, it is evident that the second condition cannot be satisfied. As a result, we can deduce that the system lacks global stability. This concern is further supported by Figure 2, which depicts the system's phase portrait.

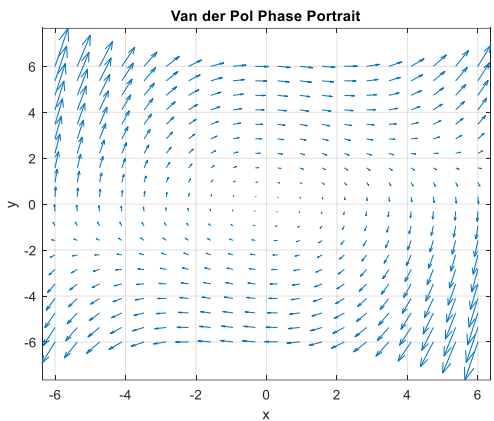

Figure 2 Van der Pol Phase portrait with stable Equilibrium point

Figure 2 illustrates that while the system possesses a locally stable equilibrium point, it cannot achieve global stability as it fails to meet the second condition required for global stability in terms of equilibrium.

## 4. Conclusion

This paper presents an innovative strategy that ensures global stability in a broad spectrum of nonlinear systems. The proposed technique expands the classical Jacobian matrix analysis by incorporating the Taylor series boundary error of estimation and the Hessian matrix eigenvalues to determine a criterion for global stability. This approach is not limited to systems with one equilibrium point or a specific dimension, which renders it superior to existing methods in global stability analysis. We demonstrate the effectiveness of our approach through several examples, including two benchmark systems. Our results show that the extended Jacobian stability analysis can ensure global stability under certain conditions, which are thoroughly discussed in this paper.

By and large, our proposed approach provides a potent tool for analyzing the global stability of nonlinear systems and has the potential to significantly advance the field of nonlinear control and optimization.

## References

[1] H. Ye, K. Zhao, H. Wu, and Y. Song, "Adaptive control with global exponential stability for parameter-varying nonlinear systems under unknown control gains," IEEE Transactions on Cybernetics, 2023.

[2] R. Iervolino and S. Manfredi, "Global stability of multi-agent systems with heterogeneous transmission and perception functions," Automatica, vol. 162, p. 111510, 2024.

[3] M. Wang, X. He, and X. Li, "Self-Triggered Impulsive Control for Lyapunov Stability of Nonlinear Systems in Discrete Time," IEEE Transactions on Cybernetics, 2024.

[4] C. M. Zagabe and A. Mauroy, "Uniform global stability of switched nonlinear systems in the Koopman operator framework," arXiv preprint arXiv:2301.05529, 2023.

[5] C. Han, Z. Liu, Y. Chen, and Q. Li, "Global Control Approach for Nonlinear Systems Contain Delayed States," in 2023 3rd International Conference on Electrical Engineering and Control Science (IC2ECS), 2023: IEEE, pp. 614-619.

[6] J. Meng, M. Yue, and D. Diallo, "Nonlinear extension of battery constrained predictive charging control with transmission of Jacobian matrix," International Journal of Electrical Power & Energy Systems, vol. 146, p. 108762, 2023.

[7] H. D. Nguyen, T. L. Vu, J.-J. Slotine, and K. Turitsyn, "Contraction analysis of nonlinear DAE systems," IEEE Transactions on Automatic Control, vol. 66, no. 1, pp. 429-436, 2020.

[8] S. Zhai and W. X. Zheng, "Stability conditions for cluster synchronization in directed networks of diffusively coupled nonlinear systems," IEEE Transactions on Circuits and Systems I: Regular Papers, vol. 70, no. 1, pp. 413-423, 2022.

[9] S. Das, D. Nandi, B. Neogi, and B. Sarkar, "Nonlinear system stability and behavioral analysis for effective implementation of artificial lower limb," Symmetry, vol. 12, no. 10, p. 1727, 2020.

[10] G. K. Rose, "Computational methods for nonlinear systems analysis with applications in mathematics and engineering," 2017